\documentclass{osa-article}
\usepackage{xcolor}
\journal{osajournal}

\usepackage{pdfpages}
\articletype{Research Article}

\begin{document}

\title{Giant Terahertz Polarization Rotation in Ultrathin Films of Aligned Carbon Nanotubes}

\author{Andrey Baydin,\authormark{1,4} Natsumi Komatsu,\authormark{1} Fuyang Tay,\authormark{1} Saunab Ghosh,\authormark{1} Takuma Makihara,\authormark{2} G.\ Timothy Noe II,\authormark{1} and Junichiro Kono\authormark{1,2,3,5}}

\address{\authormark{1}Department of Electrical and Computer Engineering, Rice University, Houston, Texas 77005, USA\\
\authormark{2}Department of Physics and Astronomy, Rice University, Houston, Texas 77005, USA\\
\authormark{3}Department of Materials Science and NanoEngineering, Rice University, Houston, Texas 77005, USA\\
\authormark{4}baydin@rice.edu\\
\authormark{5}kono@rice.edu}




\begin{abstract}
For facile manipulation of polarization states of light for applications in communications, imaging, and information processing, an efficient mechanism is desired for rotating light polarization with a minimum interaction length. Here, we report giant polarization rotations for terahertz (THz) electromagnetic waves in ultrathin ($\sim$45\,nm), high-density films of aligned carbon nanotubes. We observed polarization rotations of up to $\sim$20$^{\circ}$ and $\sim$110$^{\circ}$ for transmitted and reflected THz pulses, respectively. The amount of polarization rotation was a sensitive function of the angle between the incident THz polarization and the nanotube alignment direction, exhibiting a `magic' angle at which the total rotation through transmission and reflection becomes exactly 90$^\circ$. Our model quantitatively explains these giant rotations as a result of extremely anisotropic optical constants, demonstrating that aligned carbon nanotubes promise ultrathin, broadband, and tunable THz polarization devices. 
\end{abstract}

\section{Introduction}

Terahertz (THz) technology has made impressive advances in the last decade, finding a wide variety of applications in spectroscopy, imaging, sensing, and communications\cite{tonouchi2007}. However, basic components such as polarizers, waveplates, and filters are still limited for THz optics. For example, the widely used THz polarizers are wire-grid polarizers\cite{yeh1978}, which are fragile, inflexible, and non-adjustable; they also require precise fabrication procedures and have low extinction ratios compared to polarizers available in the infrared or visible spectral range\cite{wiesauer2013recent}. Waveplates in the THz range are usually limited to bulk crystals\cite{masson2006}. Reports on giant Faraday and Kerr rotations in thin films and crystals require stringent conditions, like applied magnetic fields and low temperatures\cite{Crassee2010, Aguilar2012, arikawa2012, arikawa2013}. 
To further advance various THz applications, robust material platforms as well as new easily implementable polarization control schemes are desired.
Realizing accessible THz technologies requires developing robust and easily implementable polarization control for this spectral region. 
Recent studies have focused on exploiting metamaterials, which enable efficient polarization rotation and conversion of THz waves\cite{wen2014, kan2015enantiomeric} with a large bandwidth\cite{zhao2018,grady2013science}, but their fabrication is typically realized based on expensive methodologies.

Here, we utilize carbon nanotubes (CNTs) for THz polarization control. CNTs are one-dimensional materials with unique photonic\cite{HarozetAl13NS, NanotetAl12AM, RenetAl12JIMT, RenetAl13PRB, ZhangetAl13NL} and optoelectronic\cite{NanotetAl13Book,AvourisetAl08NP,WeismanKono19Book} properties. Recent advances in fabricating macroscopic films of aligned single-wall CNTs\cite{HeetAl16NN} have enabled new fundamental studies and applications\cite{gao2019RS, gao2019JPD,komatsu2020groove}. For example, owing to their strong anisotropic optical properties in a broad spectral range\cite{gao2019JPD}, such films have been shown to be excellent THz polarizers\cite{ren2009nl, ren2012nl,HeetAl16NN,KomatsuetAl17AFM}, which are comparable to commercial wire-grid polarizers in terms of extinction ratio and natural hyperbolic materials in the mid-infrared\cite{GaoetAl19ACS, RobertsetAl19NL}. We observed record-high values of THz polarization rotation in ultrathin ($\sim$45~nm) films of aligned CNTs: up to $\sim$20$^{\circ}$ through single-pass transmission, and up to $\sim$110$^{\circ}$ upon a single reflection.  The amount of polarization rotation sensitively depended on the polarization angle, $\theta$, of the incident THz wave with respect to the nanotube alignment direction. At a `magic' angle ($\theta$ $\sim30^\circ$), the total rotation due to transmission plus reflection became exactly 90$^\circ$. We developed a detailed theoretical model, which quantitatively explains all experimental observations.  We demonstrate that the observed giant polarization rotations are a result of the extremely anisotropic optical constants of the CNT films, and that the magic angle can be tuned by changing the substrate refractive index and the film thickness. 
These easy-to-fabricate, robust, high-temperature resistible, ultrathin, broadband, flexible, and tunable THz polarization devices based on macroscopically aligned and densely packed CNTs will address a fundamental challenge in the development of THz optical devices. 

\begin{figure}
    \centering
    \includegraphics[width=0.7\columnwidth]{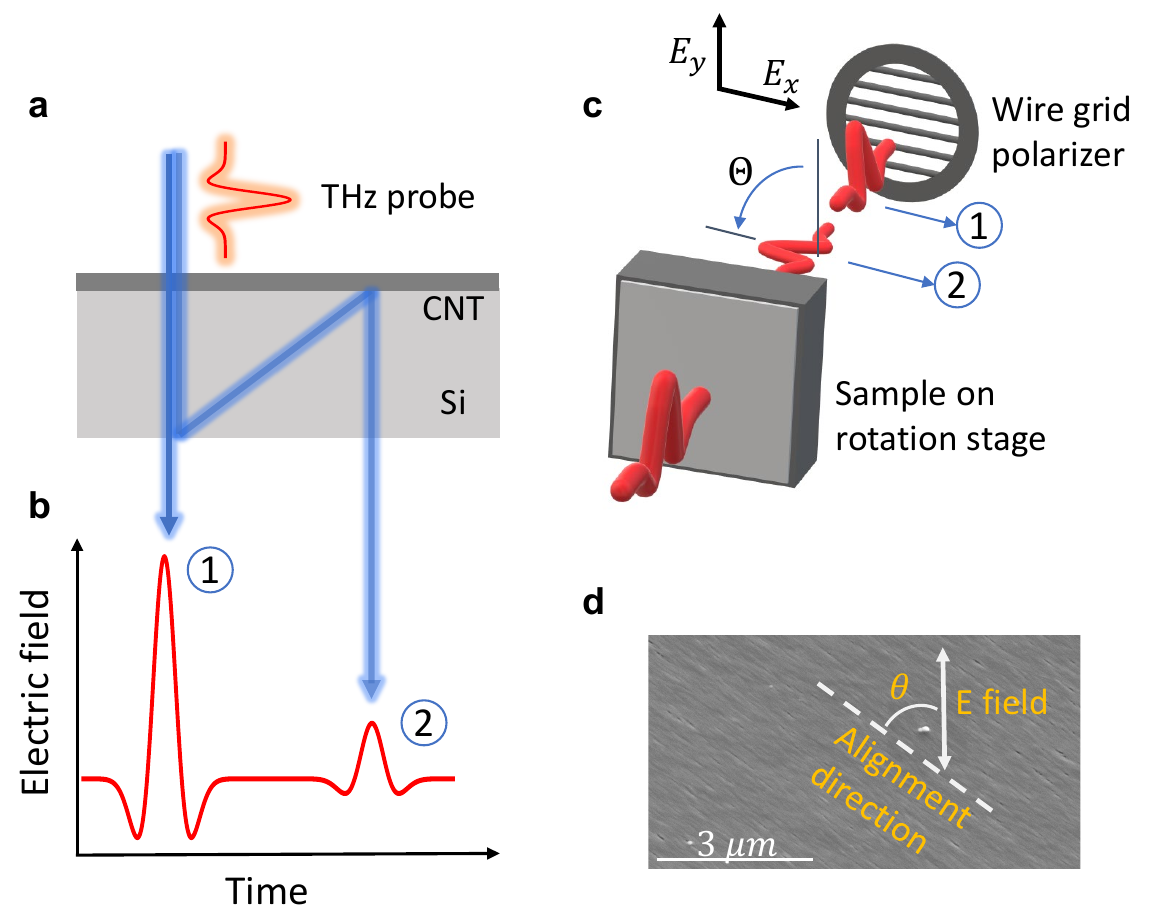}
    \caption{\textbf{Experimental setup for showing giant THz polarization rotation in an aligned CNT film.} \textbf{a},~A schematic of THz transmission and reflection through the CNT film and substrate. \textbf{b},~THz waveform in the time domain indicating the existence of a second pulse due to reflections in the substrate as shown in \textbf{a}. \textbf{c},~Experimental configuration showing wire-grid polarizer, the sample, and the schematic of the polarization rotation of the propagating THz pulse. \textbf{d},~Polarization angle $\theta$ defined as the angle between the CNT alignment direction and the polarization of the incident THz electric field.}
    \label{fig:setup}
\end{figure}

\section{Materials and Methods}
We prepared aligned single-wall CNT films using the controlled vacuum filtration method\cite{HeetAl16NN,gao2019RS,gao2019JPD,komatsu2020groove}. Arc discharge carbon nanotubes (P2-SWNT) were purchased from Carbon Solutions, Inc. The carbon nanotubes were a mixture of semiconducting and metallic species with a ratio of 2:1. Then, a dilute aqueous suspension of SWCNTs with sodium deoxycholate surfactant (0.01\%) was filtered using a vacuum filtration system at a well-controlled speed, which resulted in a wafer-scale (diameter $\sim$ 2 inches) crystalline SWCNT film in which nanotubes are nearly perfectly aligned (with nematic order parameter $S \sim 1$) and maximally packed ($\sim$1 nanotube per cross-sectional area of 1 nm$^2$). After that, the film was transferred onto a silicon substrate using a wet transfer method~\cite{HeetAl16NN}. The average length and diameter values were 300~nm and 1.4~nm, respectively.

Polarization-dependent THz time-domain spectroscopy measurements were carried out with the standard THz time-domain spectroscopy technique in a transmission geometry. THz pulses were generated via optical rectification in Mg-doped stoichiometric LiNbO$_3$ that was pumped by the output beam of an amplified Ti:Sapphire laser system (Clark-MXR, Inc., CPA-2001) producing pulses centered at 775~nm with 1~kHz repetition rate and 150~fs pulse duration. The THz beam spot diameter was estimated to be $\sim$2~mm. The THz pulses were probed in the time domain via electro-optic sampling using a ZnTe crystal. Measurements were performed inside a box purged with dry air to remove excess water vapor. The sample, an aligned SWCNT film on a substrate, was mounted on a rotation stage. When the THz pulse traverses the CNT-film/Si-substrate system, it undergoes multiple reflections -- first by the back side of the substrate and then by the substrate-CNT film interface; see Figure~1a. This results in a second pulse in the time domain, which is the focus of this paper; see Figure~1b. Only the first pulse is considered for determining the optical constants of the sample using traditional THz-TDS\cite{neu2018jap}. 

Moreover, the first pulse has been demonstrated to effectively determine the degree of alignment of aligned CNT films\cite{HeetAl16NN}. Here, we focus on the second pulse because it can be used to deduce information on the reflection properties of the CNT film. Figure~1c shows a schematic diagram of polarization rotation of THz pulses upon transmission (pulse 1) and reflection (pulse 2). The detection polarizer was used to measure the $E_x$ and $E_y$ components of the THz electric field. The experiments were performed for various incident polarization angles, $\theta$, as defined in Figure~1d.

\section{Results and Discussion}
\begin{figure}
    \centering
    \includegraphics[width=\textwidth]{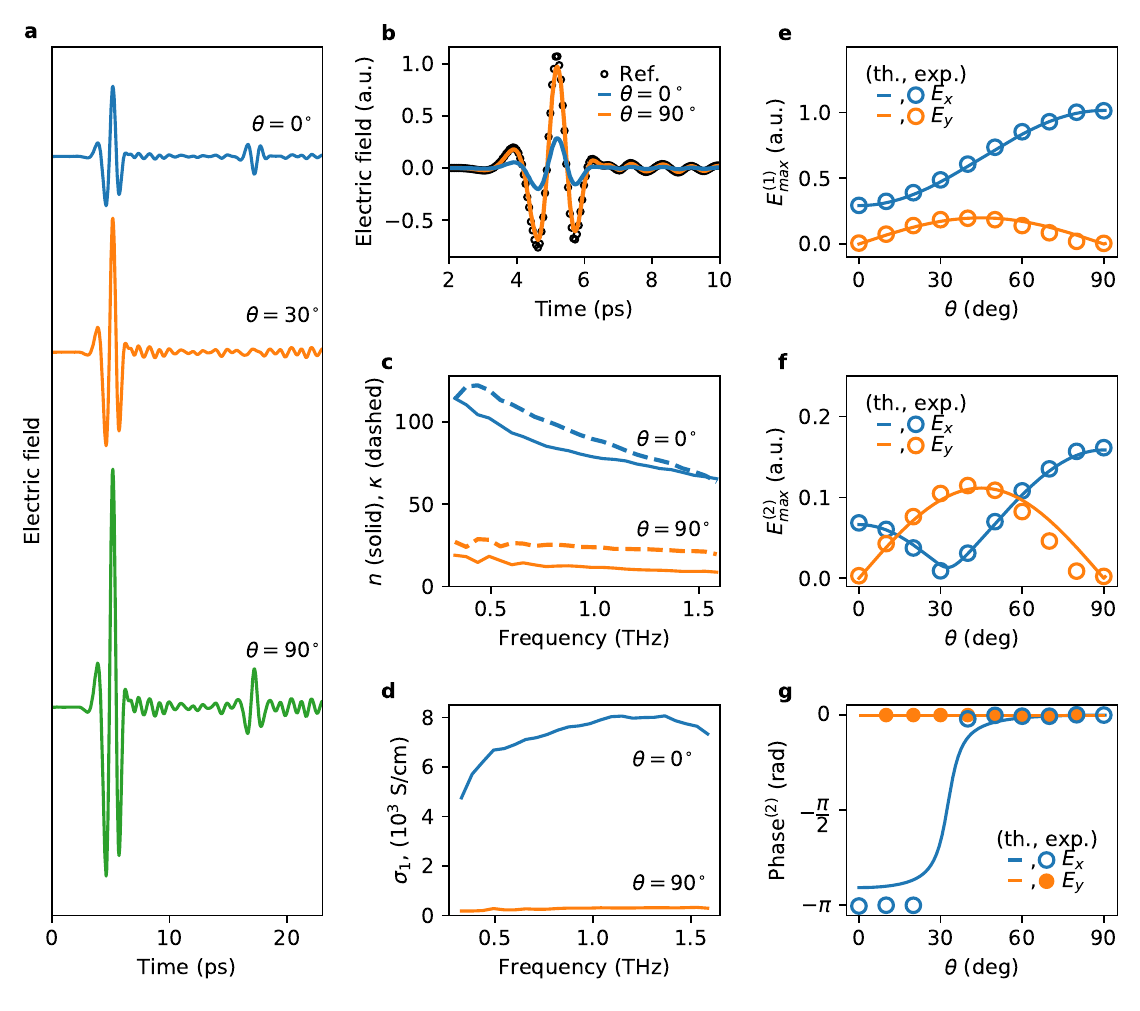}
    \caption{\textbf{Polarization angle dependence of THz signal and optical anisotropy.} \textbf{a},~The waveforms of THz pulses transmitted through the aligned SWCNT film on a Si substrate at different angles, $\theta$, between the CNT alignment direction and the incident THz field polarization direction. \textbf{b},~The time-domain waveform of the first pulse for a reference (a Si substrate), and for the CNT film for two polarization orientations with respect to the CNT alignment direction. \textbf{c},~The real (solid) and imaginary (dashed) parts of the complex refractive index and \textbf{d}~the real part of optical conductivity obtained for the CNT film. The peak amplitude of the THz electric field of the \textbf{e}~first and \textbf{f}~second pulses as a function of $\theta$. \textbf{g},~The phase of the THz electric field as a function of $\theta$. Open circles represent experimental data, and solid lines are theoretical curves. Superscript ($i = 1, 2$) indicates the pulse number. }
    \label{fig:waveforms}
\end{figure}
Figure~2a shows time-domain traces of measured THz electric fields for $\theta$ = 0$^\circ$, 30$^\circ$, and 90$^\circ$ when both the input and output polarizers were oriented in the $x$ direction (see $E_x$ in Figure~1c). The traces are vertically offset for clarity.  Both the first and second pulses are detected for any values of $\theta$, except 30$^\circ$.  The second pulse is absent only for $\theta = 30^\circ$. We refer to this angle as the `magic' angle. 
Similar disappearance of the reflection pulse has been observed and controlled by deposition of thin metal films\cite{Thoman2008}. Such a method is based on impedance matching, i.e., matching of refractive indices between a sample and a metal film. In the present case, the second pulse disappears only when the output polarizer is set to the $E_x$ orientation. When the output polarizer direction along the $E_y$ orientation, the second pulse is present except for $\theta = 0^\circ$ and $\theta = 90^\circ$. This can be clearly seen in Figures~2e and 2f, which show the $E_x$ and $E_y$ peak electric fields of the first and second THz pulses, respectively. Additional time-dependent data is shown in Fig.~S1. 

Clearly, the magic angle value depends on the anisotropic refractive index of the CNT film, which can be further tuned by doping/gating, and carbon nanotube chirality. However, these additional experiments are beyond the scope of the present paper. Instead, Figure~S2a shows the dependence of the magic angle on the underlying substrate refractive index for a fixed film thickness. The magic angle decreases as the refractive index of the substrate increases. Such dependence can be understood by the fact that by changing the refractive index of the substrate, the Fresnel transmission and reflection coefficients at the interface are modified. Furthermore, the magic angle also changes with the thickness of the CNT film, although the variation was small for the thickness range utilized (25-100~nm); see Fig.~S2b.

To understand these experimental data, let us consider the following model: a THz pulse propagates in free space (or a vacuum, ``v''), enters a system consisting of a film (``f'') of thickness $d_\mathrm{f}$ on a substrate (``s'') of thickness $d_\mathrm{s}$, experiences multiple reflections inside the system (thus producing multiple trailing pulses), and exits into free space. The complex electric field amplitude of the first ($E_1$) and second ($E_2$) THz pulses can be written as
\begin{equation}
\label{eq:efield}
    E_{1}=t_\mathrm{f} P_\mathrm{s} t_\mathrm{sv} E_\mathrm{in}
\end{equation}
\begin{equation}
    E_{2}=t_\mathrm{f} r_\mathrm{f} P_\mathrm{s}^{3} t_\mathrm{sv} E_\mathrm{in}
\end{equation}
where $E_\mathrm{in}$ is the input electric field, $t_{jk} = 2n_j/(n_j+n_k)$ and $r_{jk}=(n_j-n_k)/(n_j+n_k)$ are Fresnel transmission and reflection coefficients, respectively, and $P_j=\exp(i k_0 d_j n_j)$ is the propagation factor for the $j$-th layer. $t_\mathrm{f}$ and $r_\mathrm{f}$ are the transmission and reflection coefficients, respectively, for the CNT film in a thin-film approximation\cite{neu2018jap,Thoman2008}
\begin{equation}
\label{eq:trfilm}
\begin{aligned}
    & t_\mathrm{f} = \frac{2 n_\mathrm{s}}{n_\mathrm{s}+1+Z_{0} d \sigma} \\
    & r_\mathrm{f} = \frac{n_\mathrm{s}-1-Z_{0} d \sigma}{n_\mathrm{s}+1 + Z_{0} d \sigma}, 
\end{aligned}
\end{equation}
where $Z_0 \approx 377$\,$\Omega$ is the impedance of free space, $d$ is the film thickness, and $\sigma$ is the complex conductivity of the film. 
Using the Jones matrix formalism, the projection of the THz electric field onto the output polarizer direction as a function of $\theta$ is obtained as
\begin{equation}
\label{eq:angle}
\begin{aligned}
    & E_x^{(i)} = E^{(i)}_{\parallel} \cos^2\theta + E^{(i)}_{\perp}\sin^2\theta,\\
    & E_y^{(i)} = \left(E^{(i)}_{\parallel}-E^{(i)}_{\perp}\right)\sin\theta\cos\theta,
\end{aligned}
\end{equation}
where $E_\parallel$ and $E_\perp$ indicate the electric field components that are parallel $ (\theta=0^\circ)$ and perpendicular $(\theta=90^\circ)$ to the CNT alignment direction, respectively, and the superscript ($i$) stands for the first ($i = 1$) and second ($i = 2$) pulses. 

The complex permittivity tensor and the complex refractive index can be obtained from the first THz pulse alone, using the thin-film approximation; see Eq.~(\ref{eq:trfilm}).
Figure~2b shows THz waveforms of the first pulse for the sample and a reference. The THz pulse is significantly attenuated for parallel polarization whereas no attenuation is seen for the perpendicular case. This is a result of a highly anisotropic complex refractive index, as shown in Figure~2c. The corresponding real part of the extracted conductivity is shown in Figure~2d.

There have been many theoretical~\cite{Ando02JPSJ,Giamarchi04Book,PustilniketAl06PRL,RoschAndrei00PRL,SablikovShchamkhalova97JETP} and experimental~\cite{Nguyen2012a,UgawaetAl99PRB,BommelietAl96SSC,Hilt2000,JeonetAl02APL,RenetAl13PRB,JeonetAl04JAP,Wang2016,Wangetal2018SciRep} studies on the optical conductivity of carbon nanotubes. However, different types of carbon nanotubes (HiPco, CoMoCAT, CVD, Arc Discharge, and Laser Ablation) were used, and the degree of alignment significantly varied from study to study, and thus, universal behaviors have not been achieved as to the frequency dependence, anisotropy, and magnitude of the optical conductivity. Additionally, most of the samples experimentally studied were a mixture of semiconducting and metallic nanotubes with a wide distribution of diameters and lengths, which prevented workers from achieving universal conclusions. The main difference between our films and the films reported elsewhere is the degree of alignment and high packing density, which we achieved using the controlled vacuum filtration method~\cite{HeetAl16NN}.

By plugging the obtained conductivity into Eqs.~(\ref{eq:efield}-\ref{eq:angle}), we can calculate the magnitude and phase of the electric field for both pulses, which are shown by solid lines in Figure~2e and 2f, respectively, together with experimental data (open circles). The calculated results for a center frequency of 0.76~THz are in good agreement with the data. In contrast to the first pulse, whose amplitude monotonically varies with $\theta$, the peak amplitude of the second pulse as a function of $\theta$ has a minimum at the magic angle ($\theta = 30^\circ$) and a maximum at $\theta = 45^\circ$, as shown in Figure~2f. Another interesting observation is a $180^\circ$ or $\pi$ phase flip of the second pulse as $\theta$ is swept through the magic angle, as can be seen in Figure~2g.

The observed effects can be understood as a result of giant polarization rotation induced by transmission through and reflection from the aligned CNT film. As an incoming THz pulse of particular polarization propagates through the CNT film, its polarization plane rotates because the electric field components that are parallel, $E_\parallel$, and perpendicular, $E_\perp$, to the CNT alignment direction get attenuated and retarded differently. The parallel component, $E_\parallel$, is attenuated more than the perpendicular one, $E_\perp$, as shown in Figure~2a. A further polarization rotation occurs when the second THz pulse (produced by the reflection at the bottom surface of the substrate) gets reflected by the film-substrate interface; i.e., $E_\parallel$ is reflected more than $E_\perp$ (see the refractive index in Figure~2c). Therefore, the polarization direction of the second THz pulse can become orthogonal to the output polarizer, which results in reflection disappearance at the magic angle (see Figure~2f). Thus, upon transmission through and reflection from the ultrathin CNT film, the THz pulse polarization direction rotates by 90 degrees when $\theta$ is at the magic angle.

To assess how much the plane of polarization rotates for the  first and second pulses, we calculated the angle of polarization rotation as $\Theta = \arctan\left (T_{xy} / T_{xx} \right )$. The real part of $\Theta$ represents the rotation of the polarization plane, while the imaginary part represents the ellipticity. We refer to the angle of polarization rotation due to transmission through the CNT film as the ``Faraday'' angle, $\Theta_\mathrm{F}$, and that due to reflection from the CNT film as the ``Kerr'' angle, $\Theta_\mathrm{K}$. Note, however, that no magnetic field is applied. When calculating the angle of polarization rotation for the second pulse, $\Theta_\mathrm{F+K}$, we must subtract the contribution from the first pulse, $\Theta_\mathrm{F}$, in order to obtain $\Theta_\mathrm{K}$. Both $\Theta_\mathrm{F}$ and $\Theta_\mathrm{K}$ are shown in Figure~3 as a function of $\theta$. After THz pulse transmission, we can see that the THz pulse polarization has a rotation just over 20$^\circ$ (Figure~3a), while the second THz pulse that is reflected from the CNT film rotates by $\sim$ 110$^\circ$ when $\theta=30^\circ$. Interestingly, due to the phase flip of the second pulse (Figure~2g), the polarization rotates counter-clockwise (clockwise) when $\theta$ is below (above) the magic angle, which is depicted schematically by arrows in Figure~3b. The phase flip occurs due to the anisotropic conductivity of the film (Figure~2d). The huge conductivity for the parallel direction leads to a negative real part of the reflection coefficient [Eq.~(\ref{eq:trfilm})], while it is positive in the perpendicular case. 

\begin{figure}
    \centering
    \includegraphics[width=0.8\textwidth]{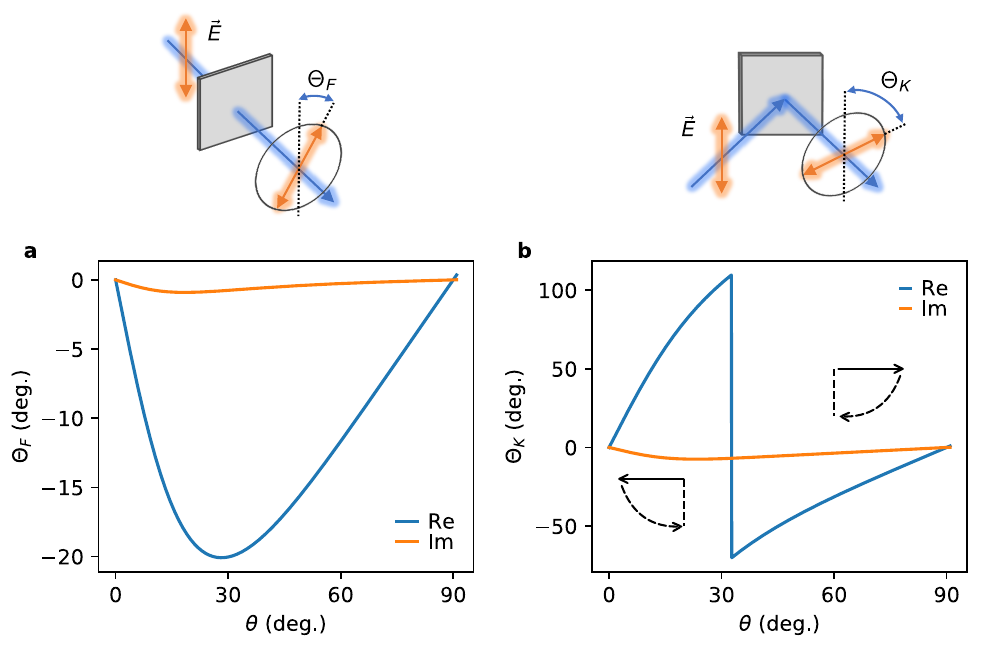}
    \caption{\textbf{Giant polarization rotations by an aligned CNT film through transmission and reflection.} \textbf{a},~The angle of polarization rotation upon transmission through the CNT film. \textbf{b},~The angle of polarization rotation upon reflection from the CNT film. Arrows in \textbf{b} indicate polarization rotation direction as the angle, $\theta$, between CNT alignment direction and input THz polarization changes. Schematics on top show definition of the polarization rotation angles.}
    \label{fig:my_label}
\end{figure}

While the magic angle depends on the refractive index of the underlying substrate, its effect on the angle of rotation is small, as can be seen in Fig. S3 and Fig. S4. It is important to note that the polarization rotation reported here arise from retardation as well as attenuation of the electromagnetic field due to the large anisotropy of both the real and imaginary parts of the refractive index with respect to the nanotube alignment direction. However, if the attenuation along one direction (i.e., for light parallel to nanotubes) is too strong, as is the case with thicker CNT films ($>$ 250~nm) or wire grid polarizers, the effect discussed in this paper will not be observed.

\section{Conclusion}
Our experiments and theoretical explanations demonstrate that, upon a single reflection by the aligned CNT film, the THz polarization plane rotated by up to $\sim$ 110 degrees. To our knowledge, this is the largest Kerr angle ever reported for a single reflection event by any material.  In contrast to other reports on giant polarization rotations based on Faraday rotation\cite{Crassee2010, Aguilar2012}, metamaterials\cite{zhao2018, wen2014,grady2013science}, and CNT functionalized gratings\cite{Xu2018}, the macroscopically aligned CNT films are broadband and ultrathin and can be put on a flexible substrate. The observed effects are explained by the extreme birefringence of the films arising from the nearly perfect alignment of CNTs. The results of this work can thus lead to CNT-based robust and flexible THz devices for manipulating THz waves.

\begin{backmatter}

\bmsection{Acknowledgments}
Authors acknowledge support from the National Science Foundation (NSF) under award no.\ ECCS-1708315, the U.S.\ Department of Energy under award no.\ DE-FG02-06ER46308, Robert A. Welch Foundation under award no.\ C-1509, and JST CREST program through Grant Number JPMJCR17I5, Japan. 

\bmsection{Disclosures}
The authors declare no conflicts of interest.

\bmsection{Data availability} Data underlying the results presented in this paper are not publicly available at this time but may be obtained from the authors upon reasonable request.

\bmsection{Supplemental document}
See Supplement 1 for supporting content. 

\end{backmatter}

\bibliography{references}



\section*{Supplementary material}

# Giant Terahertz Polarization Rotation in Ultrathin Films of Aligned Carbon Nanotubes: supplemental document

In addition to the polarization-dependent THz time-domain data for the main 45-nm thick CNT film shown in the main text, time-domain THz waveforms for another CNT film (with a thickness of 30 nm) on a Si substrate are shown in Fig. S1. The left and right panels, $E_x$(Left) and $E_y$(Right), respectively, correspond to the two orientations of the output polarizer. The input polarization was set along the $x$-axis. As the angle between the carbon nanotube alignment direction and the input light polarization changes, the pulses in the time-domain waveform change. The most drastic change, as mentioned in the main text and is the basis of this work, is the disappearance of the second pulse observed at ~30 degrees (see Fig. S1), which we refer to as the 'magic angle'.

This angle corresponds to the disappearance of the second pulse for the $E_x$ output polarization. Further, we determined how the magic angle depends on the CNT film thickness and the underlying substrate, as shown in Fig. S2. According to Fig. S2a, the magic angle changes by about 5 degrees when the CNT film thickness varies from 30 nm to 100 nm. Fig. S2b shows that increasing the refractive index of the substrate decreases the magic angle by about 10 degrees.

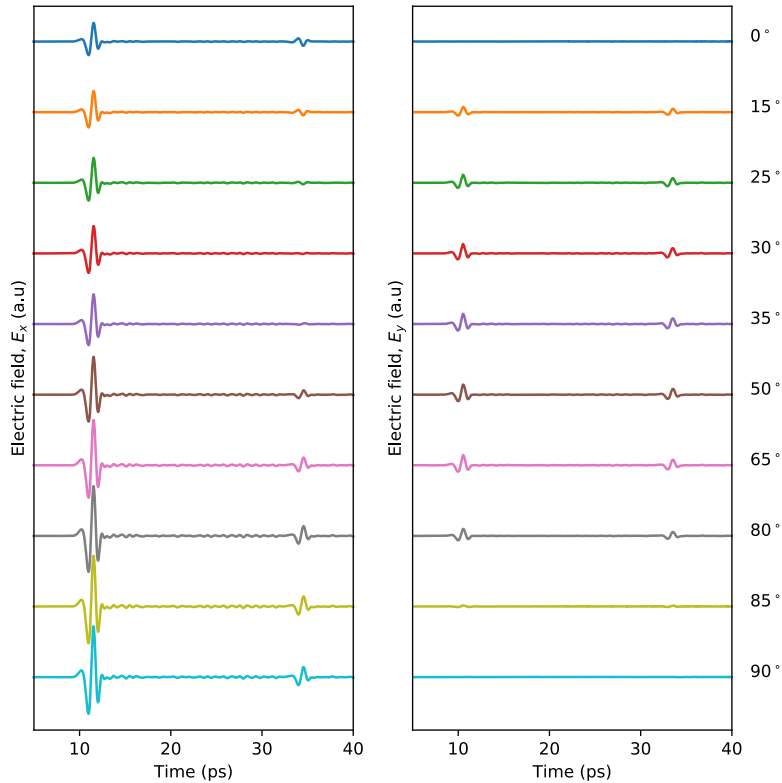

Fig. S1. Time-domain THz waveforms obtained for a 30-nm thick CNT film. The measured electric field is shown as a function of time for co-polarized, $E_x$ (Left), and counter-polarized, $E_y$ (Right), configurations. The angle between the input polarization and the alignment orientation of carbon nanotubes is indicated on the right.

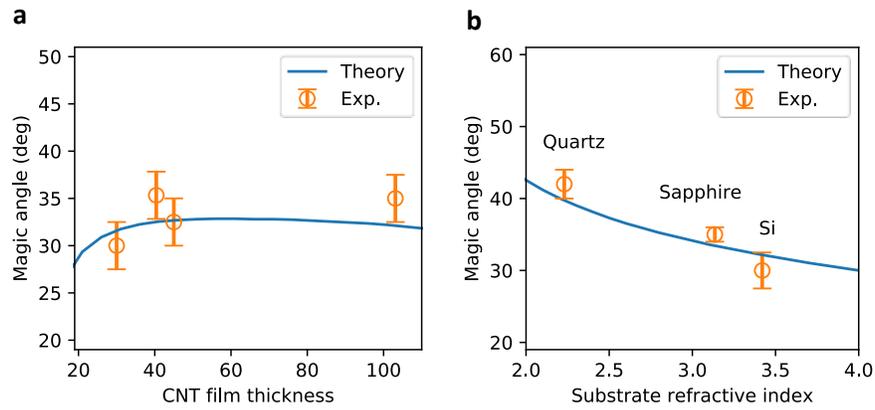

Fig. S2. The **a**, CNT film thickness dependence and the **b**, substrate dependence of the magic angle.

As discussed in the main text, the existence of the magic angle is explained by the polarization rotation of the THz wave upon transmission through and reflection by the CNT film. Fig. S3 and S4 show Faraday and Kerr angles (defined in the main text) as a function of the angle between the input light polarization and the alignment direction of carbon nanotubes for 30-nm thick CNT films on Si and quartz substrates, to complement to the data presented for the 45-nm thick CNT film on Si in the main text. In both cases, the angle of polarization rotation is about 20° and 110° for transmission and reflection, respectively.

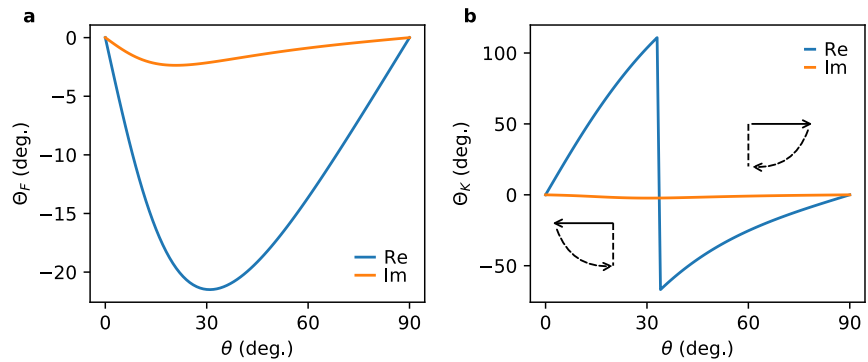

Fig. S3. Polarization rotation for the 30-nm thick CNT film on a Si substrate. **a**, The angle of polarization rotation upon transmission through the CNT film. **b**, The angle of polarization rotation upon reflection from the CNT film. Arrows indicate the polarization rotation direction as the angle, θ, between the CNT alignment direction and the input THz polarization changes.

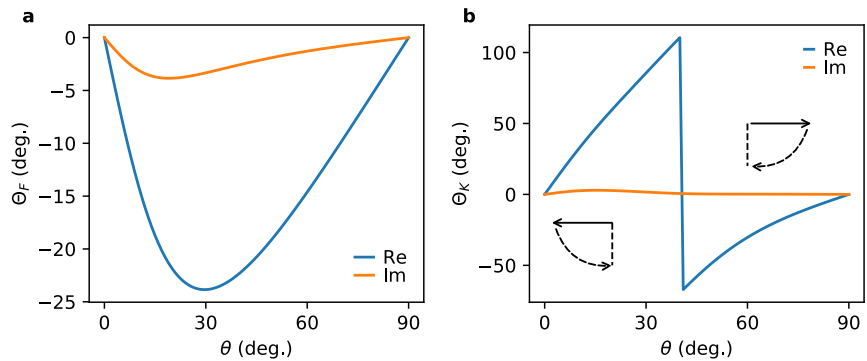

Fig. S 4. Polarization rotation for the 30-nm thick film on a quartz substrate. **a**, The angle of polarization rotation upon transmission through the CNT film. **b**, The angle of polarization rotation upon reflection from the CNT film. Arrows indicate the polarization rotation direction as the angle, θ, between the CNT alignment direction and the input THz polarization changes.


\end{document}